\documentclass[12pt]{article}
\usepackage{amsmath}
\usepackage{amssymb}
\usepackage{epsfig}
\usepackage{euscript}
\def\mathswitchr#1{\relax\ifmmode{\mathrm{#1}}\else$\mathrm{#1}$\fi}

\def\rQCED{{\rm QCED}}

\newcommand {\pslash}{\hbox{$\not\hbox{\kern-2.3pt $p$}$}}

\newcommand{\FYFS}{F_{\mathrm{YFS}}}
\usepackage{cite}

\usepackage{epic}
\def\alf1{ {\alpha\over\pi} }
\def\rQCED{{\rm QCED}}
\begin{document}
%\input{feynman} 
%=======================================================================
\begin{titlepage}
\begin{flushleft}
%{\bf CERN-PH-TH/2011-077}\\
{\bf BU-HEPP-14-10}\\
{\bf Dec., 2014}\\
\end{flushleft}
%\vspace{0.05cm}
%\begin{titlepage}
\begin{center}
{\bf \large Interplay between IR-Improved DGLAP-CS Theory and the Precision of an NLO ME Matched Parton Shower MC in Relation to LHCb Data}\\
%\end{center}
%\ShortTitle{Exact Amplitude-Based Resummation in Quantum Field Theory: Recent Results}
\vspace{2mm}
%\begin{center}
%%  {\bf   S. Jadach$^{a,b}$ and B.F.L. Ward$^{c,d}$}
%\author{B.F.L. Ward\\%
%    \thanks{Work supported in part by D.o.E. grant DE-FG02-09ER41600.}\\
%      Baylor University\\
%        E-mail:\email{bfl\_ward@baylor.edu}}
%    S.K. Majhi
% \footnote{Work supported by 
%grant Pool No. 8545-A, CSIR, IN.}\\
%      Indian Association for the Cultivation of Science, Kolkata, India\\
%        E-mail: tpskm@iacs.res.in\\
A. Mukhopadhyay\\%
%    \thanks{Work supported in part by D.o.E. grant DE-FG02-09ER41600.}\\
      Baylor University, Waco, TX, USA\\
        E-mail: aditi\_mukhopadhyay@baylor.edu\\
B.F.L. Ward\\%
%    \thanks{Work supported in part by D.o.E. grant DE-FG02-09ER41600.}\\
      Baylor University, Waco, TX, USA\\
        E-mail: bfl\_ward@baylor.edu\\
%S.A. Yost
% \footnote{Work supported in part by U.S.
%D.o.E. grant DE-FG02-10ER41694 and grants from The Citadel Foundation.}\\
%      The Citadel, Charleston, SC, USA\\
%        E-mail: scott.yost@citadel.edu\\
\end{center}
\vspace{2mm}
\centerline{\bf Abstract}
We use comparison with recent LHCb data on single $Z/\gamma^*$ production and decay to lepton pairs as a vehicle to study the current status of the application of our approach of {\it exact} amplitude-based resummation in quantum field theory to precision QCD calculations, by realistic MC event generator methods, as needed for precision LHC physics. This represents an extension of the phase space of our previous studies based on comparison with CMS and ATLAS data, as the pseudo-rapidity range measured by the LHCb  for leptons in the data we study is $2.0<\eta< 4.5$ to be compared with $|\eta|<4.6(2.4)$ in our previous CMS(ATLAS) data comparison for the same processes. The analyses we present here with the LHCb data thus represent an important addition to our previous results, as it is essential that theoretical predictions be able to control all of the measured phase space at LHC. The level of agreement between the new theory and the data continues to be a reason for optimism. \\
\vspace{1cm}
%\begin{center}
% BU-HEPP-14-ww, Aug., 2014\\
%\end{center}

%          BU-HEPP-11-04,\\ 
%             Dec., 2011 }
%\end{center}
\end{titlepage}
%\begin{document}
%\\
%\vskip 3mm
%\centerline{Invited talk presented by B.F.L. Ward at RADCOR 2011, Chennai, India}
%\vskip 16mm
% 
%\vspace{10mm}
%\renewcommand{\baselinestretch}{0.1}
%\footnoterule
%\noindent
%{\footnotesize
%\begin{itemize}
%\item[${\dagger}$]
%Work partly supported by US DOE grant DE-FG02-09ER41600. 
% the Polish Government
%grants KBN 2P30225206 and 2P03B17210, the Maria Sk\l{}odowska-Curie
%Joint Fund II PAA/DOE-97-316, and
%by NATO Grant PST.CLG.980342.
%, and by
%Polish Government grant 5P03B09320.
%\end{itemize}
%}
%\vspace{0.5cm}
%\begin{flushleft}
%{\bf UTHEP-00-0101}\\
%{\bf Jan, 2000}\\
%\end{flushleft}

%\end{titlepage}
 
%\baselineskip=11pt 
%=======================================================================
\def\Kmax{K_{\rm max}}\def\ieps{{i\epsilon}}\def\rQCD{{\rm QCD}}
%\renewcommand{\theequation}{\arabic{equation}}
%\font\fortssbx=cmssbx10 scaled \magstep2
%\renewcommand\thepage{}
%\vfill\eject
%\parskip.1truein\parindent=20pt\pagenumbering{arabic}\par

\section{\bf Introduction}\par

With the recent discovery~\cite{atlas-cms-2012} 
of a Brout-Englert-Higgs (BEH)~\cite{EBH}  
boson after the start-up and successful running 
of the LHC for three years, as we investigate its properties
and any hints in the LHC data for physics beyond the Standard Model,
we realize that we have entered the era of precision QCD, 
by which we mean
predictions for QCD processes at the total precision tag of $1\%$ or better.
We have argued in Refs.~\cite{radcor2011,radcor2013,herwiri,1305-0023,qced} that 
exact, amplitude-based 
resummation of large higher order effects is a viable strategy 
to achieve such precision tags and we have developed the MC HERWIRI1.031
~\cite{herwiri} as a first platform in this connection, with some emphasis on its MC@NLO/HERWIRI1.031 realization of exact NLO matrix-element matched parton shower calculus. 
We have previously illustrated the comparison of the new MC, which carries the 
IR-improved~\cite{irdglap1,irdglap2} DGLAP-CS~\cite{dglap,cs} theory
in a HERWIG6510~\cite{hwg} environment, with the data of ATLAS~\cite{atlaspt} and CMS~\cite{cmsrap} at the LHC and with the data of D0~\cite{d0pt} and CDF~\cite{galea} at FNAL. In what follows, we extend these studies to the data of LHCb~\cite{lhcbdata} at the LHC.\par
More precisely, any precision theory platform should be able to cover the entire
observable phase space for hard processes in order to fully exploit the data at the LHC. In the single $Z/\gamma^*$ production and decay to lepton pairs, the LHCb probes the regime in the lepton pseudo-rapidity 
$\eta= -\ln(\tan(\theta/2))$ given by 
$2.0<\eta<4.5$, where $\theta$ is the polar angle with respect 
to a beam direction. This should be compared to the regimes probed by the data we studied in Refs.~\cite{herwiri,1305-0023} which for ATLAS had $|\eta|<2.4$ for $e^+,\; e^-$ and which for CMS had $|\eta|<2.1$ for $\mu^+,\;\mu^-$ and $|\eta|<4.6$
for $e^+,\; e^-$. The cuts on the lepton transverse momenta were similar with $p_T>20$ GeV/c for all the data we discuss here. Thus, the LHCb data provide a different check on the comparison between the theoretical predictions, where we are interested in comparing the IR-improved results with NLO exact ME/shower matching to the corresponding unimproved ones, both with and without the 'ad hocly' hard intrinsic ${\rm PTRMS}\cong 2.2$ GeV/c that we have found was necessary for the unimproved calculations to explain both the rapidity and the $p_T$ spectra from ATLAS and CMS that we discussed in Refs.~\cite{herwiri}. Here, ${\rm PTRMS}$ is 
the rms value of a Gaussian intrinsic $p_T$ distribution 
for the proton constituents
in HERWIG65~\cite{hwg}. \par
What we are particularly interested to see is how the IR-improvement interplays with the change in the phase space for the accepted lepton pairs in the $Z$ mass region. In all of our work, we rely on the data as given by the experimentalists so that all conclusions we draw must be interpreted with this understanding: 
{\it our new IR-improved MC has not been used, as yet, by any of the LHC collaborations in unfolding their data in any way in obtaining their publicly available results, as far as we know, so that this limits the strength of our conclusions in a direct way}. For example, if a bin-to-bin migration effect has been estimated with a parton shower MC that has an ad hoc infrared cut-off, it is unknown what the same effect would be if it were estimated with our new IR-improved MC which does not need such a cut-off. Here, we want to try to put some emphasis on the following point. In the usual parton shower with the ad hoc infrared cutoff, let us call it $k_0$, the region of the shower emission phase space below the energy $k_0$ is dropped; for, the exponentiating virtual correction, $V_{sh}$, is defined to be the negative of the integral of the shower's real emission distribution $R$ so that when one sums over all real emissions one gets the total correction factor $e^{V_{sh}+\int R}=e^{0}=1$ and the shower does not change the normalization. Moreover, as the $k_0$ dependence in the integral over the real emission in the shower exactly cancels against that in the virtual correction, one formally has no dependence on $k_0$. But, this is only formally true. In reality, the actual events the shower makes depend on $k_0$, with softer values generically generating more soft radiation than harder values, so comparison with events in the experiments can illustrate very well the effects of changing $k_0$. In the IR-improved patron shower MC HERWIRI1.031, which realizes IR-improved DGLAP-CS theory~\cite{dglap,cs} in the 
HERWIG6.5~\cite{hwg} environment, the regime below $k_0$ is taken into account by amplitude based resummation methods as presented in Refs.~\cite{irdglap1,irdglap2}. This is expected to produce events that more closely resemble those seen in the experiments as we have illustrated in Refs.~\cite{herwiri,1305-0023}.\par
More specifically, we discuss in what follows the comparisons between the IR-improved and unimproved parton shower(PS) MC predictions, with
the exact MC@NLO~\cite{mcatnlo} PS/ME matched exact ${\cal O}(\alpha_s)$ correction, and the LHCb data on the $Z/\gamma*$ rapidity, $p_T$ and $\phi^*_\eta$ distributions, where the variable 
$\phi^*_\eta$ was introduced in Refs.~\cite{phietastr} as an attempt to overcome some of the apparent experimental difficulties in measuring finely binned $p_T$ spectra for lower values of $p_T$ for the $Z/\gamma^*$ in the LHC environment.
We have the definition $\phi_\eta^*=\tan(\frac{1}{2}(\pi-\Delta\phi))\sin\theta^* \cong \left|\sum \frac{{p_i}_T \sin\phi_i}{Q}\right| +{\cal O}(\frac{{{p_i}_T}^2}{Q^2})$, where $\Delta\phi=\phi_1-\phi_2$ is the azimuthal angle 
between the two leptons which have transverse momenta $\vec{p_i}_T,\; i=1,2,$
and $\theta^*$ is the scattering angle of the dilepton system relative to the beam direction when one boosts to the frame along the beam direction such that the leptons are back to back. We present here the comparisons with  
the respective MC@NLO~\cite{mcatnlo} parton shower/matrix element matched exact ${\cal O}(\alpha_s)$ corrections included, as this correction is now established to be important~\cite{herwiri}.\par
Our discussion proceeds as follows. In the next section, we give a brief review
of the theoretical paradigm we are using here, as this approach to precision QCD is still not a familiar one. In the Sect. 3, we turn to comparison with the LHCb data; we also discuss attendant theoretical implications therein.
In Sect. 4 we sum up.
\section{Review of Exact Amplitude-Based Resummation Theory and Its Parton Shower MC Implementation}

Our discussion here starts from the following
fully differential representation of a hard LHC scattering process:
\begin{equation}
d\sigma =\sum_{i,j}\int dx_1dx_2F_i(x_1)F_j(x_2)d\hat\sigma_{\text{res}}(x_1x_2s),
%       &=\sum_{i,j}\int dx_1dx_2{F'}_i(x_1){F'}_j(x_2)\hat\sigma'(x_1x_2s),
%\end{split}
\label{bscfrla}
\end{equation}
where the $\{F_j\}$ and 
$d\hat\sigma_{\text{res}}$ are the respective parton densities and 
resummed reduced hard differential cross section which 
has been resummed
for all large EW and QCD higher order corrections in a manner consistent
with achieving a total precision tag of 1\% or better for the  
theoretical precision of (\ref{bscfrla}). As we have explained
in Refs.~\cite{herwiri,1305-0023}, to have 
the latter precision tag as a realistic goal, 
we have developed 
the $\text{QCD}\otimes\text{QED}$ resummation theory in Refs.~\cite{qced}
for the reduced cross section in (\ref{bscfrla}) and for the
resummation of the evolution of the parton densities therein as well.
\par
Specifically, 
for both the resummation of the reduced cross section
and that of the evolution of the parton densities,
the defining formula
may be identified as
% In (\ref{bscfrla}), resummation 
%of collinear evolution is realized in the evolution of the $\{F_j\}$ and 
%soft resummation(non-collinear) is realized in the calculation of 
%$d\hat\sigma_{\text{res}}$. For example, from Ref.~\cite{qced} we have the 
%representation
\begin{eqnarray}
&d\bar\sigma_{\rm res} = e^{\rm SUM_{IR}(QCED)}
   \sum_{{n,m}=0}^\infty\frac{1}{n!m!}\int\prod_{j_1=1}^n\frac{d^3k_{j_1}}{k_{j_1}} \cr
&\prod_{j_2=1}^m\frac{d^3{k'}_{j_2}}{{k'}_{j_2}}
\int\frac{d^4y}{(2\pi)^4}e^{iy\cdot(p_1+q_1-p_2-q_2-\sum k_{j_1}-\sum {k'}_{j_2})+
D_\rQCED} \cr
&\tilde{\bar\beta}_{n,m}(k_1,\ldots,k_n;k'_1,\ldots,k'_m)\frac{d^3p_2}{p_2^{\,0}}\frac{d^3q_2}{q_2^{\,0}},
%\end{split}
\label{subp15b}
\end{eqnarray}\noindent
where $d\bar\sigma_{\rm res}$ is either the reduced cross section
$d\hat\sigma_{\rm res}$ or the differential rate associated to a
DGLAP-CS~\cite{dglap,cs} kernel involved in the evolution of the $\{F_j\}$ and 
where the {\em new} (YFS-style~\cite{yfs,yfs-jw}) {\em non-Abelian} residuals 
$\tilde{\bar\beta}_{n,m}(k_1,\ldots,k_n;k'_1,\ldots,k'_m)$ have $n$ hard gluons and $m$ hard photons and we show the final state with two hard final
partons with momenta $p_2,\; q_2$ specified for a generic $2f$ final state for
definiteness. The infrared functions ${\rm SUM_{IR}(QCED)},\; D_\rQCED\; $
are defined in Refs.~\cite{qced,irdglap1,irdglap2} as follows:
\begin{eqnarray}
{\rm SUM_{IR}(QCED)}=2\alpha_s\Re B^{nls}_{QCED}+2\alpha_s{\tilde B}^{nls}_{QCED}\cr
D_\rQCED=\int \frac{d^3k}{k^0}\left(e^{-iky}-\theta(K_{max}-k^0)\right){\tilde S}^{nls}_{QCED}
\label{irfns}
\end{eqnarray}
where the dummy parameter $K_{max}$ is such that nothing depends on it and where we have introduced
\begin{eqnarray}
B^{nls}_{QCED} \equiv B^{nls}_{QCD}+\frac{\alpha}{\alpha_s}B^{nls}_{QED},\cr
{\tilde B}^{nls}_{QCED}\equiv {\tilde B}^{nls}_{QCD}+\frac{\alpha}{\alpha_s}{\tilde B}^{nls}_{QED}, \cr
{\tilde S}^{nls}_{QCED}\equiv {\tilde S}^{nls}_{QCD}+{\tilde S}^{nls}_{QED}.
\label{irfns1}
\end{eqnarray} 
Here, the superscript $nls$ denotes that the infrared functions are DGLAP-CS synthesized as explained in Refs.~\cite{dglpsyn,qced,irdglap1,irdglap2} and the infrared functions
$B_A,\; {\tilde B}_A,\; {\tilde S}_A, \; A=QCD,\; QED,$ are given
in Refs.~\cite{yfs,yfs-jw,qced,irdglap1,irdglap2}\footnote{We note that the ratio of QED and QCD couplants on the RHS of (\ref{irfns1}) is suppressed in Ref.~\cite{1305-0023}, for example.}. 
The  
simultaneous resummation of QED and QCD large IR effects is exact here.
See Refs.~\cite{herwiri,1305-0023} for discussion of the physical meanings of the various components of the master formula as well as for discussion
of our connection with the methods in Ref.~\cite{gatheral}. Here, we do continue to stress that
in our formulation in (\ref{subp15b})
{\it the entire soft gluon phase space is included in the 
representation -- no part of it 
is dropped}.
\par
The new non-Abelian residuals $\tilde{\bar\beta}_{m,n}$ 
allow~\cite{qced} rigorous shower/ME matching via their shower subtracted analogs. What this means is that
in (\ref{subp15b}) we make the replacements
\begin{equation}
\tilde{\bar\beta}_{n,m}\rightarrow \hat{\tilde{\bar\beta}}_{n,m}
\end{equation}
where the $\hat{\tilde{\bar\beta}}_{n,m}$ have had all effects in the showers
associated to the $\{F_j\}$ removed from them. Contact between the $\hat{\tilde{\bar\beta}}_{n,m}$ and the
differential distributions in MC@NLO proceeds as
follows. Representing the no-emission probability, the Sudakov from factor, as
$$\Delta_{MC}(p_T)=e^{[-\int d\Phi_R \frac{R_{MC}(\Phi_B,\Phi_R)}{B}\theta(k_T(\Phi_B,\Phi_R)-p_T)]},$$ 
the MC@NLO differential cross section can be written as~\cite{mcatnlo} 
\begin{equation}
\begin{split}
d\sigma_{MC@NLO}&=\left[B+V+\int(R_{MC}-C)d\Phi_R\right]d\Phi_B[\Delta_{MC}(0)+\int(R_{MC}/B)\Delta_{MC}(k_T)d\Phi_R]\\
&\qquad\qquad +(R-R_{MC})\Delta_{MC}(k_T)d\Phi_Bd\Phi_R
\label{mcatnlo1}
\end{split}
\end{equation}
where $B$ is Born distribution, $V$ is the regularized virtual contribution,
$C$ is the corresponding counter-term required at exact NLO, $R$ is the respective
exact real emission distribution for exact NLO and  $R_{MC}=R_{MC}(P_{AB})$ is the parton shower real emission distribution, with the obvious notation for the respective phase spaces $\{d\Phi_A, \; A=B,\; R\}$.
From comparison with (\ref{subp15b}) restricted to its QCD aspect we get~\cite{herwiri,1305-0023} the identifications, accurate to ${\cal O}(\alpha_s)$,
\begin{equation}
\begin{split}
\frac{1}{2}\hat{\tilde{\bar\beta}}_{0,0}&= \bar{B}+(\bar{B}/\Delta_{MC}(0))\int(R_{MC}/B)\Delta_{MC}(k_T)d\Phi_R\\
\frac{1}{2}\hat{\tilde{\bar\beta}}_{1,0}&= R-R_{MC}-B\tilde{S}_{QCD}
\label{eq-mcnlo}
\end{split}
\end{equation}
where we defined~\cite{mcatnlo} $$\bar{B}=B(1-2\alpha_s\Re{B_{QCD}})+V+\int(R_{MC}-C)d\Phi_R$$ and we understand here
that the DGLAP-CS kernels in $R_{MC}$ are to be taken as the IR-improved ones
as we exhibit below~\cite{irdglap1,irdglap2}. 
Here the QCD virtual and real infrared functions
$B_{QCD}$ and $\tilde{S}_{QCD}$ are understood to be DGLAP-CS synthesized as explained in Refs.~\cite{qced,irdglap1,irdglap2} to
avoid double counting of effects. In view of 
(\ref{eq-mcnlo}), 
the way to the extension of frameworks such as MC@NLO to exact higher
orders in $\{\alpha_s,\;\alpha\}$ is therefore open via our $\hat{\tilde{\bar\beta}}_{n,m}$
and will be taken up elsewhere~\cite{elswh,1407-7290}.
\par
In Ref.~\cite{1305-0023} we have presented  detailed discussion of the relationship between our approach to precision QCD and those presented in Refs.~\cite{stercattrent1,scet1,colsop,colsopster}. We refer the reader interested in this relationship to
Ref.~\cite{1305-0023}. Here, we focus on the observation that strict control on the theoretical precision
in (\ref{bscfrla}) requires both the resummation of the reduced cross section
and that of the attendant evolution of the $\{F_j\}$. We turn now
to the latter. 
\par 
More precisely, we get an improvement
of the IR limit of the kernels, $P_{AB}$, in the DGLAP-CS theory itself
when we apply the QCD restriction of the formula in (\ref{subp15b}) to the
calculation of these kernels. This is the IR-improved DGLAP-CS 
theory~\cite{irdglap1,irdglap2} in which large IR effects are resummed 
for the kernels themselves.
The attendant new resummed kernels, $P^{\exp}_{AB}$, whose 
implementation in the HERWIG6.5 environment generates the new 
MC HERWIRI1.031~\cite{herwiri},
are as follows~\cite{irdglap1,irdglap2,herwiri}:
{\small
\begin{align}
P^{\exp}_{qq}(z)&= C_F \FYFS(\gamma_q)e^{\frac{1}{2}\delta_q}\left[\frac{1+z^2}{1-z}(1-z)^{\gamma_q} -f_q(\gamma_q)\delta(1-z)\right],\nonumber\\
P^{\exp}_{Gq}(z)&= C_F \FYFS(\gamma_q)e^{\frac{1}{2}\delta_q}\frac{1+(1-z)^2}{z} z^{\gamma_q},\nonumber\\
P^{\exp}_{GG}(z)&= 2C_G \FYFS(\gamma_G)e^{\frac{1}{2}\delta_G}\{ \frac{1-z}{z}z^{\gamma_G}+\frac{z}{1-z}(1-z)^{\gamma_G}\nonumber\\
&\qquad +\frac{1}{2}(z^{1+\gamma_G}(1-z)+z(1-z)^{1+\gamma_G}) - f_G(\gamma_G) \delta(1-z)\},\nonumber\\
P^{\exp}_{qG}(z)&= \FYFS(\gamma_G)e^{\frac{1}{2}\delta_G}\frac{1}{2}\{ z^2(1-z)^{\gamma_G}+(1-z)^2z^{\gamma_G}\},
%P_{qG}(z)&=\frac{1}{2}(z^2+(1-z)^2).
\label{dglap19}
\end{align}}
where the superscript ``$\exp$'' indicates that the kernel has been resummed as
we described. 
Here
$C_F$($C_G$) is the quadratic Casimir invariant for the quark(gluon) color representation respectively, and  
the YFS~\cite{yfs} infrared factor 
is given by $$\FYFS(a)=e^{-C_Ea}/\Gamma(1+a)$$ where $\Gamma(w)$($C_E=0.57721566...$) is Euler's gamma function(constant), respectively. 
The respective resummation functions $\gamma_A,\delta_A,f_A, A=q,G$ are given in Refs.~\cite{irdglap1,irdglap2}
\footnote{The improvement in Eq.\ (\ref{dglap19}) 
should be distinguished from the 
resummation in parton density evolution for the ``$z\rightarrow 0$'' 
Regge regime -- see for example Refs.~\cite{ermlv,guido}. This
latter improvement must also be taken into account 
for precision LHC predictions.}
\footnote{We follow the development of Field in Ref.~\cite{rfield} in removing the mass singularities such that our evolution variable for the DGLAP-CS equation is $t=\ln(Q^2/\Lambda^2)$ as it is given in eq.(4.6.11) in Ref.~\cite{rfield}.}. These new kernels provide us with a new IR-improved resummed scheme for the parton density functions (PDF's) and the reduced cross section with the same value of $\sigma$ in (\ref{bscfrla}): 
\begin{equation}
\begin{split}
F_j,\; \hat\sigma &\rightarrow F'_j,\; \hat\sigma'\; \text{for}\\
P_{Gq}(z)&\rightarrow P^{\exp}_{Gq}(z), \text{etc.}
\end{split}
\label{newscheme1}
\end{equation}
As discussed in Refs.~\cite{herwiri}, this new scheme has improved MC stability
-- in the attendant parton shower MC HERWIRI1.031 based on the new kernels there is no need 
for an IR cut-off `$k_0$'
parameter. While the degrees of freedom
below the IR cut-offs in the usual showers are dropped in those showers,
in the showers in HERWIRI1.031, these degrees of freedom are included in the calculation and are integrated over in the process of generating the Gribov-Lipatov exponents $\gamma_A$ in (\ref{dglap19}). We note also that, {\em as the differences between them start in ${\cal O}(\alpha_s^2)$, the new kernels
agree with the usual kernels at ${\cal O}(\alpha_s)$}. Thus, 
for the realization of exact
NLO ME/shower matching the MC@NLO and POWHEG frameworks apply directly 
to the new kernels. 
\par
In Fig.~1, in the interest of pedagogy, the basic physical idea,
discussed by Bloch and Nordsieck in Ref.~\cite{bn1}, which underlies the new kernels is illustrated -- we show this because our approach is still not generally a familiar one: 
\begin{figure}[h]
\begin{center}
\epsfig{file=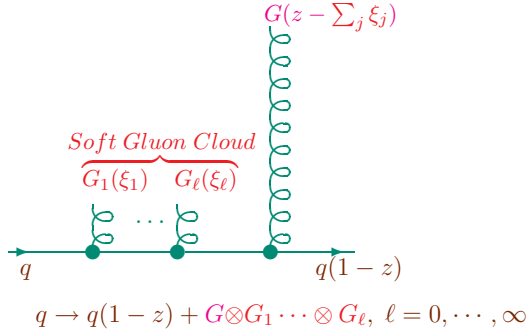,width=70mm}
\end{center}
\label{fig-bn-1}
\caption{Bloch-Nordsieck soft quanta for an accelerated charge.}
\end{figure}
a coherent state of very soft massless quanta of the attendant gauge field is 
generated by an accelerated charge so that one cannot know which of the 
infinity of possible states
has been made in the splitting process $q(1)\rightarrow q(1-z)+G\otimes G_1\cdots\otimes G_\ell,\; \ell=0,\cdots,\infty$ shown in Fig.~1.
This effect is taken into account in the new kernels 
by resumming the 
terms ${\cal O}\left((\alpha_s \ln(\frac{q^2}{\Lambda^2})\ln(1-z))^n\right)$
in the $z\rightarrow 1$ IR limit.  
We see from (\ref{newscheme1}) and (\ref{bscfrla}) 
that when the usual kernels are used these terms
are generated order-by-order in the solution for the cross section
$\sigma$ in (\ref{bscfrla}). Thus, for a given order of exactness in the
input perturbative components therein our 
resumming them enhances the convergence of the 
representation in (\ref{bscfrla}).  
This last remark is illustrated in the next Section in the context of the comparison of recent LHCb data to
NLO parton shower/matrix element matched predictions.\par

\section{Interplay of IR-Improved DGLAP-CS Theory and NLO Shower/ME Precision: Comparison with LHCb Data}

In the new MC HERWIRI1.031~\cite{herwiri} we have the first realization of the new IR-improved kernels in the HERWIG6.5~\cite{hwg} environment. In Refs.~\cite{herwiri,1305-0023}, we have made comparisons with the data of ATLAS and CMS on the single $Z/\gamma*$ production and decay to lepton pairs and have been encouraged by the results of these comparisons. Here, we extend our comparisons to the more forward acceptance of the LHCb for the same processes.\par
We need to make the following important observation regarding the results
from Refs.~\cite{herwiri,1305-0023}. We have shown that the exact ${\cal O}(\alpha_s)$ correction makes a $11.5\%$ -- $12.8\%$ correction to the spectra 
in $p_T$ and $Y$ analyzed therein. Following the methodology in Ref.~\cite{jad-prec}, we then estimate the attendant physical theoretical precision error
due to uncalculated higher orders as one-half of this effect, which we then double to be on the safe side. Thus, we will take the conservative estimate 
$\Delta\sigma^{phys}_{th}/\sigma\cong 13\%$ for the physical theoretical precision in the plots which make herein as to be understood in interpreting the discussion which we give in what follows. This will reflect what we have learned from Refs.~\cite{herwiri,1305-0023} regarding the errors in our work on the theoretical side. This estimate should be considered preliminary and serves as a lower bound; we expect our final value to be somewhat larger. We would, however, like to emphasize that what we do here represents
a somewhat different approach to the error estimate than what is done in Ref.~\cite{fewz2}, for example, wherein the various scales of the calculation are varied by some arbitrarily chosen factors. The authors in Ref.~\cite{blm} have emphasized the probable limitations of such approaches. Our approach is based 
on our experiences in the precision LEP EW studies~\cite{jad-prec} and 
more generically on the analyses of Refs.~\cite{zinn-justin}. 
As a practical matter,
we will omit our physical theoretical precision error estimate from the $\chi^2/\text{d.o.f.}$ which we use to
compare theory and data, so that our $\chi^2/\text{d.o.f.}$ are respective upper limits accordingly.\par
\begin{figure}[h]
\begin{center}
%x\epsfig{file=pent-1.eps,width=140mm}
%\includegraphics[width=100mm]{fig-doe-2012-2a.eps}
\setlength{\unitlength}{0.1mm}
\begin{picture}(1600, 930)
\put( 400, 770){\makebox(0,0)[cb]{\bf (a)} }
\put(1240, 770){\makebox(0,0)[cb]{\bf (b)} }
\put(   -50, 0){\makebox(0,0)[lb]{\includegraphics[width=90mm]{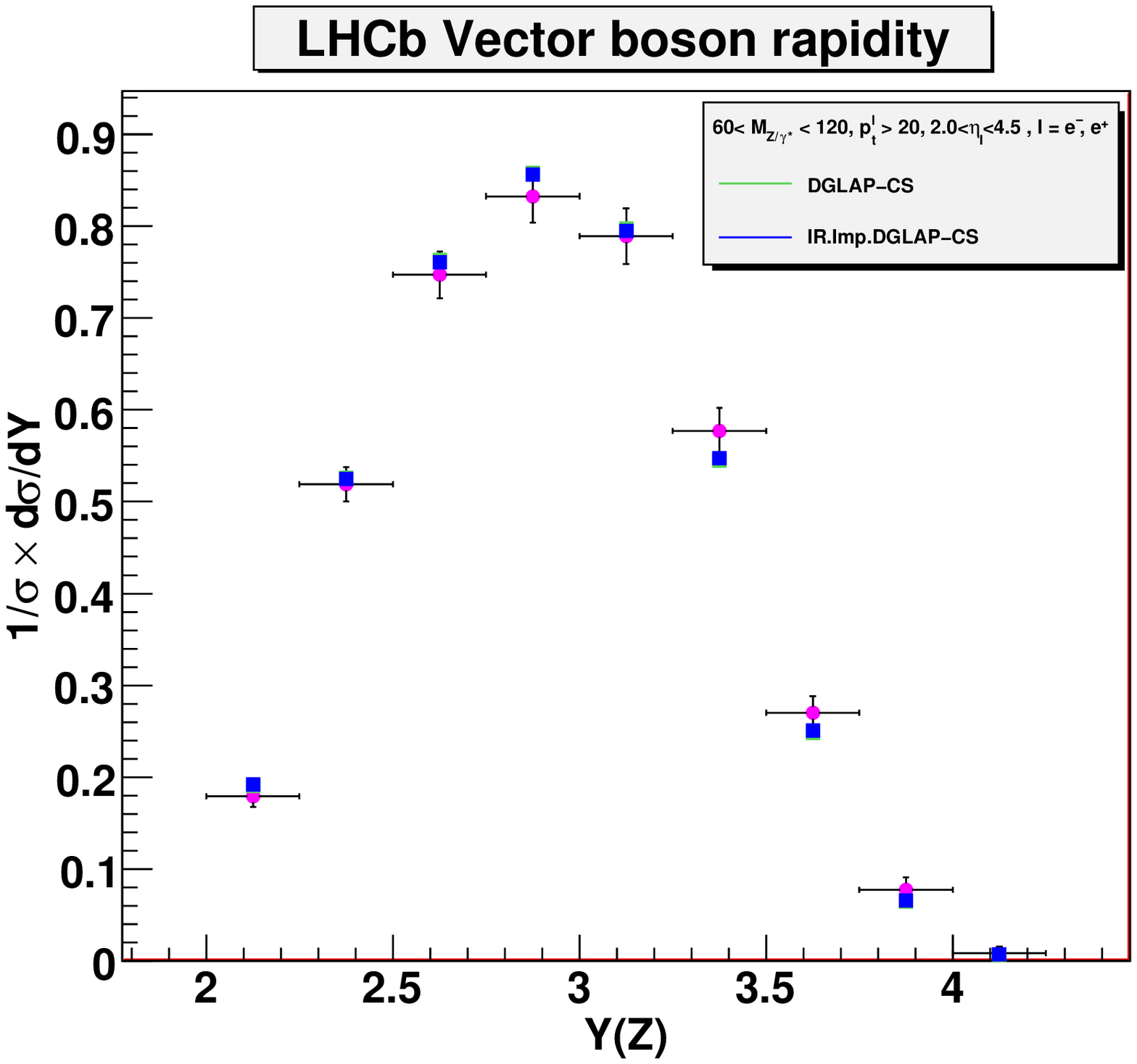}}}
\put( 800, 0){\makebox(0,0)[lb]{\includegraphics[width=90mm]{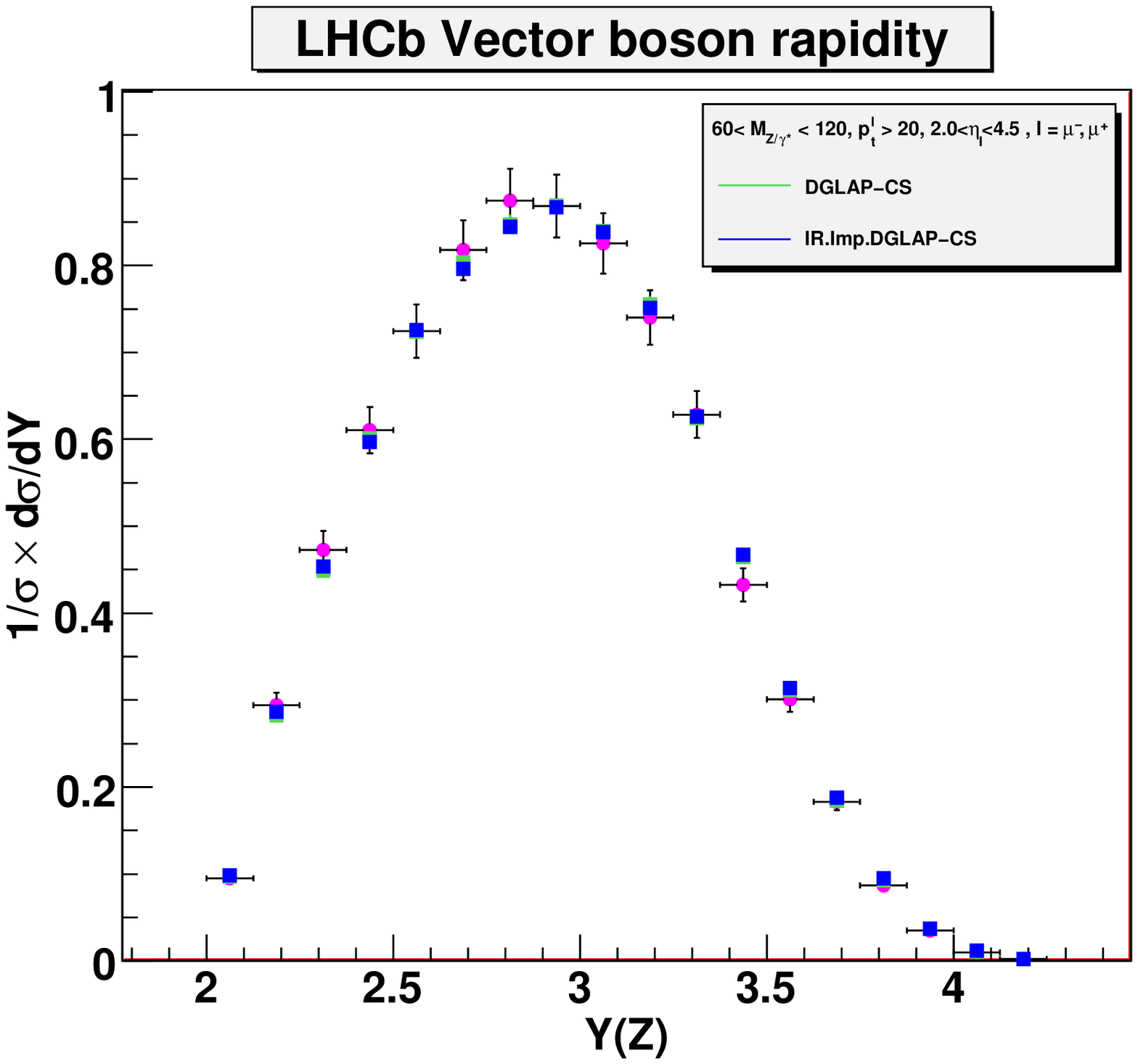}}}
\end{picture}
\end{center}
\caption{\baselineskip=8pt Comparison with LHCb data: (a), LHCb rapidity data on
($Z/\gamma^*$) production to $e^+e^-$ pairs, the circular dots are the data, the green(blue) squares are MC@NLO/HERWIG6.510($\rm{PTRMS}=2.2$ GeV/c)(MC@NLO/HERWIRI1.031); 
(b), LHCb rapidity data on ($Z/\gamma^*$) production to (bare) $\mu^+\mu^-$ pairs, with the same graphical notation as that in (a). In both (a) and (b), the green triangles are MC@NLO/HERWIG6.510($\rm{PTRMS}=$0). These are otherwise untuned theoretical results. 
}
\label{figlhcb1}
\end{figure}

    
Specifically, with the recent LHCb data~\cite{lhcbdata} as our baseline,
we compare the predictions from HERWIRI1.031 with HERWIG6.510, with
the MC@NLO~\cite{mcatnlo} exact ${\cal O}(\alpha_s)$ correction included,
to illustrate the interplay between the attendant precision in NLO ME matched parton shower MC's  
and the new IR-improvement for the kernels.
In Fig.~\ref{figlhcb1} in panel (a) we show for the single $Z/\gamma*$ production at the LHC 
the comparison between the LHCb rapidity data for the $e^+e^-$ channel
and the MC theory predictions and in panel
(b) in the same figure we show the analogous comparison with the LHCb data
for the $\mu^+\mu^-$ channel.  
These results should be considered from the perspectives of our analysis in Refs.~\cite{herwiri} of the FNAL data on the single $Z/\gamma^*$ production in 
$\text{p}\bar{\text{p}}$ collisions at $1.96$ TeV and our analysis in Ref.~\cite{herwiri,1305-0023} of the LHC ATLAS and CMS data single $Z/\gamma^*$ production in 
$\text{p}\text{p}$ collisions. More precisely, what we found in Refs.~\cite{herwiri,1305-0023} was that the IR-improvement in HERWIRI1.031 allowed it to give a better $\chi^2/\text{d.o.f}$ to the FNAL and ATLAS and CMS data without the need
of a large intrinsic value $\text{PTRMS}$ on the scale of the expectations from successful models of the proton~\cite{pwvfn}, which would have $\text{PTRMS}\simeq 0.4$ GeV/c, as would the precociousness of Bjorken scaling~\cite{bj1,scaling}. This is in contrast to the unimproved results from HERWIG6.5, wherein such a large value as $2.2$ GeV/c for $\text{PTRMS}$ was needed to get similar values of
 $\chi^2/\text{d.o.f}$ for the $p_T$ spectra but for the rapidity data
such a large value of $\text{PTRMS}$
was not necessary to get the corresponding similar 
values. What we see in Fig.~\ref{figlhcb1} is that the situation is similar
in the LHCb rapidity data. The values of the $\chi^2/\text{d.o.f}$
are 0.746, 0.814, 0.836 for the respective predictions from
MC@NLO/HERWIRI1.031, MC@NLO/HERWIG6.5($\text{PTRMS}=0$) and
MC@NLO/HERWIG6.5($\text{PTRMS}=2.2$ GeV/c) for the $e^+e^-$ data and are
0.773, 0.555, 0.537 for the respective predictions for the $\mu^+\mu^-$ data.
All three calculations give acceptable values of $\chi^2/\text{d.o.f}$.\par
\begin{figure}[h]
\begin{center}
%x\epsfig{file=pent-1.eps,width=140mm}
\includegraphics[width=100mm]{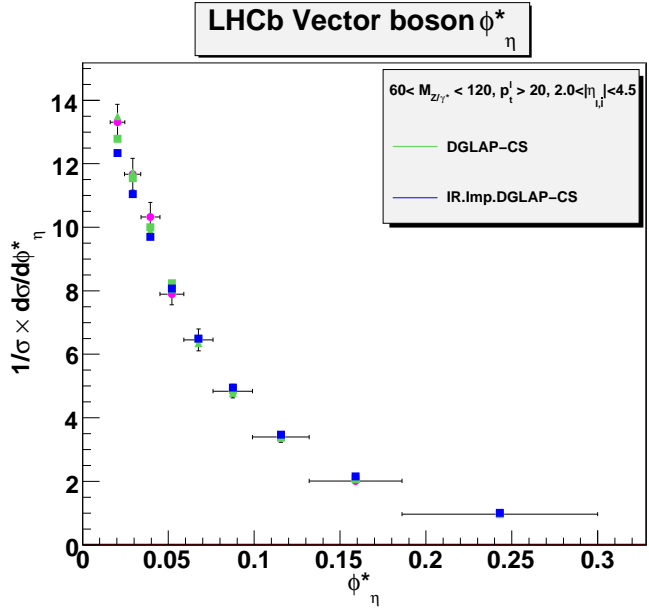}
\end{center}
\caption{\baselineskip=8pt Comparison with LHCb data on $\phi_\eta^*$ for the $\mu^+\mu^-$ channel in single $Z/\gamma^*$ production at the LHC. The legend (notation) for the plots is the same as in Fig.~\ref{figlhcb1}. 
}
\label{figlhcb2}
\end{figure}

When turn to the transverse momentum degrees of freedom, the situation changes
relative to what we found in Refs.~\cite{herwiri,1305-0023} for D0, CDF, ATLAS and CMS. We start with the $\phi_\eta^*$ LHCb data in Ref.~\cite{lhcbdata}. Specifically, we have 
\begin{equation}
\phi_\eta^*=\tan(\phi_{\text{acop}}/2)\sqrt{1-\tanh^2(\Delta\eta/2)}
\label{eqn-phi-eta*}
\end{equation}
where $\Delta\eta=\eta^--\eta^+$ when $\eta^-$ and $\eta^+$ are the respective
negatively and positively charged lepton pseudo-rapidities and $\phi_{\text{acop}}=\pi - \Delta\phi$ with $\Delta\phi$ as defined in Sect. 1. This variable is not exactly the same as the $p_T$ of the produced $Z/\gamma^*$ but is correlated with it, where as shown in Ref.~\cite{atlas-phi-star} there is a longer and longer tail in the correlation as $\phi_\eta^*$ decreases toward smaller values.
Moreover, there is significant bin-to-bin migration in the lower bins of
the $\phi_\eta^*$ data as it has been unfolded in Ref.~\cite{lhcbdata}. Thus, 
unless we have the same MC unfolding effect for each MC which we use in our comparisons, we cannot address the uncertainty of the comparison where the migration is substantial. Accordingly, we restrict the comparison to the regime where the unfolding effect is minimal until such a time when each MC will have been used for the unfolding corrections. Again, the comparisons we show here are otherwise untuned comparisons. And, we only show the MC@NLO/A results, 
for A = HERWIG6.5($\text{PTRMS}=0$), HERWIG6.5($\text{PTRMS}=2.2$ GeV/c) and HERWIRI1.031, where 
as usual we always set $\text{PTRMS}=0$ in HERWIRI1.031 simulations.
The corresponding results are shown in Fig.~\ref{figlhcb2}.
The respective $\chi^2/\text{d.o.f.}$ are 1.2, 0.23, 0.35 for the MC@NLO/ERWIRI1.031, MC@NLO/HERWIG6.5($\text{PTRMS}=0$), MC@NLO/HERWIG6.5($\text{PTRMS}=2.2$ GeV/c) simulations. Thus, all three simulations give acceptable fits to the data, with the curious result that the MC@NLO/HERWIG6.5\\
($\text{PTRMS}=0$) gives a very mildly better fit than does 
MC@NLO/HERWIG6.5($\text{PTRMS}=2.2$ GeV/c) -- we would caution here 
already that we really do not have the errors under such control that 
we could take $|\Delta\chi^2/\text{d.o.f.}|\simeq 0.1$ as significant. 
These results make us recall the difference between the $\phi_\eta^*$ 
variable and the $p_T$ of the $Z/\gamma^*$ as well as, perhaps, 
the difference between the forward
and more central observations, as
we have seen in Refs.~\cite{herwiri,1305-0023} that a good $p_T$-fit for the central region with HERWIG6.5 is not possible with $\text{PTRMS}=0$. We thus look next at the more forward LHCb data on $p_T$.
\par
In analyzing the LHCb $p_T$ data in Ref.~\cite{lhcbdata}, we note that there is significant bin-to-bin migration at low $p_T$ in the first bin. Since the different MC's have significantly different predictions in the low $p_T$ regime below $4$ GeV/c, until these effects are determined with each MC, it is not appropriate in our approach to try  to use the data to assess the accuracy of the MC's in this regime. What we will do again here is to use the data where these type of effects are minimal and wait until
we can get data wherein each MC has been used to assess these migration effects before we try to assess the regime below $p_T \cong 4$ GeV/c. We call this our ``conservative'' approach to the data.\par
\begin{figure}[h]
\begin{center}
%x\epsfig{file=pent-1.eps,width=140mm}
\includegraphics[width=100mm]{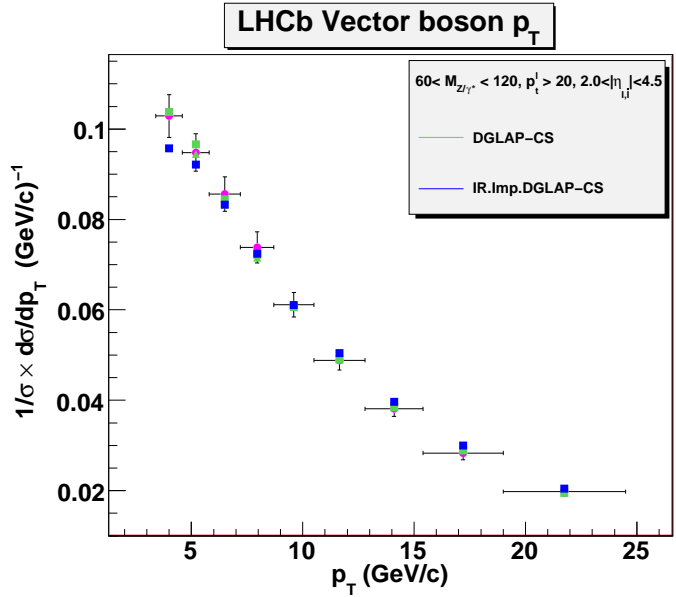}
\end{center}
\caption{\baselineskip=8pt Comparison with LHCb data on $p_T$ for the $\mu^+\mu^-$ channel in single $Z/\gamma^*$ production at the LHC. The legend (notation) for the plots is the same as in Fig.~\ref{figlhcb1}. 
}
\label{figlhcb3}
\end{figure}

With this understanding, we show in Fig.~\ref{figlhcb3} the comparisons between the three\\
MC@NLO/A predictions and the LHCb $p_T$ data on single $Z/\gamma^*$ production and decay to $\mu^+\mu^-$ pairs, where, again, A = HERWIG6.5($\text{PTRMS}=0$), HERWIG6.5($\text{PTRMS}=2.2$ GeV/c) and HERWIRI1.031, and 
as usual we always set $\text{PTRMS}=0$ in HERWIRI1.031 simulations.
What we find is that the respective $\chi^2/\text{d.o.f.}$ are 0.183, 0.103, 0.789 respectively. We see that for the more forward LHCb data all three calculations give acceptable fits to the data with only a very mild indication that
the $\text{\rm PTRMS}=2.2$ GeV/c HERWIG6.5 results give a better fit than do the
$\text{\rm PTRMS}=0$ GeV/c HERWIG6.5 results, in contrast with what was found
in Refs.~\cite{herwiri,1305-0023} for the D0 and ATLAS $p_T$ data and in general
agreement with our results in Fig.~\ref{figlhcb2} for the $\phi_\eta^*$ variable when we recall again that we really do not have the errors under such control that we can consider $|\Delta\chi^2/\text{d.o.f.}|\simeq 0.1$ as significant. We conclude that, when we look over the data on single $Z/\gamma^*$ production at FNAL(CDF and D0) and at LHC(ATLAS, CMS and LHCb), as we have shown in Refs.~\cite{herwiri,1305-0023} and with the results presented here, HERWIRI1.031 gives a good fit to all the data analyzed without the need of ad hocly hard intrinsic $\text{\rm PTRMS}$ whereas HERWIG6.5 needs such a value at $\sim 2$ GeV/c in order to give a good fit to all of these data for both the rapidity and the transverse momentum based observables. This gives us a well-defined starting point from which to set a baseline~\cite{1407-7290} for a rigorous treatment
of the theoretical precision tag on such processes at the LHC and at FCC~\cite{fcc}.\par 
\section{Conclusions}
What we have shown is the following. The realization of IR-improved DGLAP-CS theory 
in HERWIRI1.031, when used in the MC@NLO/HERWIRI1.031 exact ${\cal O}(\alpha_s)$ ME matched parton shower framework,
affords one the opportunity to explain, on an event-by-event basis, both the rapidity and the $p_T$ dependent spectra of the $Z/\gamma^*$ in pp collisions
in the recent LHC data from the LHCb, respectively, without the need of an
unexpectedly hard intrinsic Gaussian $p_T$ distribution with rms value of $\rm{PTRMS}\cong 2$ GeV/c in the proton's wave function. This extends a similar conclusion to the LHCb that we had established in Refs.~\cite{herwiri,1305-0023} for the ATLAS and CMS data. Our view is that this can be interpreted as providing 
further support for a rigorous basis for the phenomenological correctness 
of such unexpectedly hard distributions insofar as describing these data using the usual unimproved DGLAP-CS showers is concerned. Accordingly, we 
continue to propose 
that comparison of other distributions such as the invariant mass distribution 
with the appropriate cuts and the more detailed $Z/\gamma^*$ $p_T$ 
spectra in the regime below $10.0$GeV/c be used to
differentiate between these phenomenological 
representations of parton shower physics
in MC@NLO/HERWIG6.510 and the fundamental description of the parton shower physics in MC@NLO/HERWIRI1.031. We recall that elsewhere~\cite{1305-0023} we have further emphasized that the precociousness of Bjorken scaling~\cite{scaling,bj1} argues against the fundamental correctness 
of the {\em hard} scale intrinsic $p_T$ ansatz with the unexpectedly large value of $\rm{PTRMS}\cong 2$ GeV/c, as do the successful models~\cite{pwvfn} of the proton's wave function,
which would predict this value to be $\lesssim 0.4$ GeV/c. As we have emphasized as well elsewhere~\cite{1305-0023},
we point out that the fundamental description in MC@NLO/HERWIRI1.031 can be systematically improved to the NNLO parton shower/ME matched level~\cite{znder-hoeche} -- a level which we anticipate is a key ingredient in achieving the (sub-)1\% precision tag for such processes as single heavy gauge boson production at the LHC.
Our comparisons with the LHCb data are not inconsistent the proposition that our methods should work in its region of the acceptance phase space as well as they do in the 
acceptance phase spaces for ATLAS and CMS. \par 
In closing, one of us (B.F.L.W.)
thanks Prof. Ignatios Antoniadis for the support and kind 
hospitality of the CERN TH Unit while part of this work was completed.\par

\end{document}